\documentclass[pre,twocolumn,aps,superscriptaddress]{revtex4-1}
\usepackage [dvips]{graphicx}
\usepackage{amsmath}
\usepackage{mathrsfs}
\usepackage{color}

\newcommand{\bomega}{\boldsymbol{\omega}}
\newcommand{\bGamma}{\boldsymbol{\Gamma}}
\newcommand{\blambda}{\boldsymbol{\lambda}}
\newcommand{\beeta}{\boldsymbol{\eta}}
\newcommand{\bpi}{\boldsymbol{\pi}}
\newcommand{\dd}{\text{d}}

\newcommand{\ee}{\text{e}}

\newcommand{\p}{\partial}
\newcommand{\bx}{\text{\bf x}}
\newcommand{\br}{\text{\bf r}}

\newcommand{\eps}{\varepsilon}
\newcommand{\bk}{\text{\bf k}}

\newcommand{\bp}{\text{\bf p}}

\newcommand{\bh}{{\bf h}}
\newcommand{\bH}{{\bf H}}
\newcommand{\bv}{\text{\bf v}}

\newcommand{\bPi}{\boldsymbol{\Pi}}

\newcommand{\bmu}{\boldsymbol{\mu}}
\newcommand{\bmm}{\text{\bf m}}

\newcommand{\comments}[1]{}   

\begin{document}

\title{Stochastic dynamics of collective modes for Brownian dipoles}
\author{Leticia F. Cugliandolo}
\affiliation{Sorbonne Universit\'es, Universit\'e Pierre et Marie Curie - Paris 6 \\
Laboratoire de Physique Th\'eorique et Hautes \'Energies, UMR 7589 CNRS/P6,
4 place Jussieu, 75252 Paris cedex 05, France}
\author{Pierre-Michel D\'ejardin}
\affiliation{Laboratoire de Math\'ematiques et de Physique, Universit\'e de Perpignan Via Domitia, 52 avenue Paul Alduy, 66860 Perpignan cedex, France}
\author{Gustavo S. Lozano}
\affiliation{Departmento de F\'isica,
Facultad de Ciencias Exactas y Naturales, Universidad de Buenos Aires
Ciudad Universitaria, Pabellon I, 1428 Buenos Aires,
Argentina}
\author{Fr\'ed\'eric van Wijland}
\affiliation{Laboratoire Mati\`ere et Syst\`emes Complexes, UMR 7057 CNRS/P7,
Universit\'e Paris Diderot, 10 rue Alice Domon et L\'eonie
Duquet, 75205 Paris cedex 13, France}
\affiliation{Department of Chemistry, University of California, Berkeley,CA94720, USA}

\begin{abstract}
The individual motion of a colloidal particle is described by an overdamped Langevin equation. When rotational degrees of freedom are relevant, these are
described by a corresponding Langevin process. Our purpose is to show that the microscopic local density of colloids,
in terms of a space and rotation state, also evolves according to a Langevin equation. The latter can then be used as the starting point of a variety of
approaches, ranging from dynamical density functional theory to mode-coupling approximations.

\end{abstract}

\maketitle

\section{The interest of studying collective modes}
Back in 1994, Kawasaki~\cite{Kawasaki199435} proposed a phenomenological equation for the evolution of the local density modes of an assembly of interacting 
colloids, the individual dynamics of which are governed by an overdamped Langevin process. Kawasaki's evolution equation was then used as a starting point in dynamical 
density functional studies of supercooled liquids. However, beyond its phenomenological nature, the exact meaning of the coarse-grained density that the Kawasaki equation 
theory deals with was unclear. A clarifying work by Dean~\cite{0305-4470-29-24-001} came a little later: an exact Langevin equation bearing on the microscopic colloidal 
density was put forward. Dean's equation has the pleasant feature that it involves the free-energy of a density profile in a physically transparent form (with 
an entropic term arising from the ideal gas contribution, and the standard two-body interaction for the potential energy term). Beyond soon-to-be-resolved 
controversies~\cite{jcp.124.16.10.1063.1.2189243,0953-8984-12-8A-356, 0305-4470-33-15-101} that have to do with the singular nature of the local density field 
(this is formally a sum of Dirac delta's centered around each particle's position), the idea of Dean was then extended~\cite{1751-8121-42-6-065001} to underdamped 
Langevin dynamics, yielding consistency with momentum-conserving fluctuating hydrodynamics, and the relation to Dynamical Density Functional Theory (DDFT)  was 
further discussed by Yoshimori~\cite{PhysRevE.71.031203}. The statistical properties of the Brownian gas were recovered in~\cite{1751-8121-41-23-235002}
with an exact perturbative analysis of Dean's equation.
Non-trivial concrete applications (that we are aware of) pertain to the 
realm of glassy dynamics. The equation was exploited by a series of authors, Miyazaki and Reichman~\cite{0305-4470-38-20-L03}, Andreanov, Biroli and 
Lef\`evre~\cite{1742-5468-2006-07-P07008}, Basu and Ramaswamy~\cite{1742-5468-2007-11-P11003}, Kim and 
Kawasaki~\cite{1751-8121-40-1-F04, 1742-5468-2008-02-P02004}, and more recently by Kim, Kawasaki, Jacquin and Van 
Wijland~\cite{PhysRevE.89.012150}. Connections of Dean's derivation
with other field-theory based approaches were discussed by Velenich, Chamon, Cugliandolo and Kreimer~\cite{1751-8121-41-23-235002} and Andreanov, 
Biroli, Bouchaud, and Lef\`evre~\cite{PhysRevE.74.030101}. The motivation behind this series of works is to provide a description of the slowing down of the 
dynamics in the ``supercooled" 
regime of colloidal glass formers directly in terms of collective modes, rather than resorting to the cumbersome projection operator technique. This is easier said 
than done, but it is a developing research direction at present.

When colloids further possess rotational degrees of freedom, a number of physical complications emerge, as discussed by Han {\it et al.}~\cite{Han27102006}, 
in a careful study of the rotational Brownian motion of ellipsoidal colloids. The assumed one-to-one correspondence~\cite{PhysRevLett.101.028101} between 
rotational and translational diffusion is thus questioned, especially in active matter studies where it is often important to filter out thermal fluctuations from the 
contributions of the active processes driving the system of interest out of equilibrium~\cite{PhysRevE.85.011905}. Similar questions arise at the level of 
collective behavior, most notably in the field of glasses, where translational and rotational degrees of freedom have been shown to respond differently to 
temperature or density changes. DDFT including rotational degrees of freedom was
derived and analyzed by Bagchi and Chandra~\cite{BagchiChandra} and Wittkowski and L\"owen~\cite{doi:10.1080/00268976.2011.609145}, in the spirit 
of the early approach of Kawasaki~\cite{Kawasaki199435}.

Our goal in this work is to establish a stochastic evolution equation for the microscopic local density of rigid dipoles, with given orientation and position. 
We will show that it takes the form of a Langevin equation with some features that are very similar to the ones discussed by Dean~\cite{0305-4470-29-24-001}, 
but with a number of differences due to the more complex individual rotational dynamics. We shall begin with a brief description of the rotational 
motion of a single colloid in terms of a Langevin process. We will then construct a Langevin equation for the local density of a set of interacting 
particles. Our conclusion will point to a number of direct applications that we plan for the future.

\section{Dynamics of a single particle}

The translational motion of a particle $i$ is characterized by its time-dependent position $\br_i$ and velocity ${\mathbf v}_i$. 
Each particle carries an, also time-dependent, electric dipole orientation ${\bf p}_i$. We assume for simplicity that the molecules are not polarizable so that their dipolar momentum is simply $\bp_i$.
Although constant in modulus, the orientation of the dipole is time dependent and can be characterized by an angular velocity vector
$\bomega_i$ such that $\frac{\dd \bp_i}{\dd t}=\bomega_i\times\bp_i$. It is subjected to an external force 
${\bf F}_i$ and to an external torque $\bGamma_i$. For a particle of mass $m$ and inertia tensor $I$ (we have thin rods in mind), we have that
\begin{eqnarray}
\label{eqdebase}
m\frac{\dd\bv_i}{\dd t}&=&-\gamma\bv_i+{\bf F}_i+\beeta_i
\; ,
\\
\label{eqdebase1}
I\frac{\dd\bomega_i}{\dd t } &=&-\zeta\bomega_i+\bGamma_i+\blambda_i
\; .
\end{eqnarray}
Here $\beeta_i$ and $\blambda_i$ are Gaussian random forces and torques, respectively, introduced to account for the thermal exchanges with the surrounding medium. The friction coefficients $\gamma$ and $\zeta$ govern the dissipation into the thermal bath. Just as $\gamma$ and $\beeta_i$ are related by a Stokes-Einstein relation, a similar relation between $\zeta$ and $\blambda_i$ exists. We refer to the existing literature~\cite{debye,McConnell1977,doi:10.1142/9789814355674} for a comprehensive presentation of the motivations behind the modeling in \eqref{eqdebase}-\eqref{eqdebase1}
which we take for granted. As explained in these references, the Gaussian random contributions $\beeta_i$ and $\blambda_i$ have $\delta$-correlations in time, the amplitude of which is constrained by the condition that for conservative forces the equilibrium distribution should be the standard Boltzmann-Gibbs exponential factor,
$\langle \eta_i(t) \eta_j(t') \rangle = 2\gamma k_BT \delta_{ij} \delta(t-t')$, and similarly for 
$\lambda_i(t)$ with $\gamma$ replaced by $\zeta$. 
Details about this can be found in \cite{doi:10.1142/9789814355674}. In physical conditions under which inertial effects can
be discarded, at low Reynolds numbers, we obtain a set of overdamped Langevin equations
\begin{eqnarray}
\label{Langevin}
\gamma\frac{\dd \br_i}{\dd t} &=& {\bf F}_i+\beeta_i
\; ,
\\
\label{Langevin1}
\zeta\frac{\dd \bp_i}{\dd t} &=& \bGamma_i\times\bp_i+\blambda_i\times {\bp_i}
\; ,
\end{eqnarray}
in which the latter equation, which features a multiplicative noise, is to be understood with the Stratonovich, mid-point, discretization scheme. 
Indeed, the Stratonovich convention is manifestly consistent with the conservation of the modulus of the dipole, as can be seen directly from 
\eqref{Langevin1}, but we will not make explicit use of that property  in the course of our derivation (though it is of course duly preserved).
The dangers and subtleties of this Langevin equation have been recently thoroughly investigated in \cite{discretizationgilbert}. The ingredients entering the force ${\bf F}_i$ felt by particle $i$ include an external force field and possible interactions with other colloids (which, for simplicity, we will assume to be two-body). Similarly, the torque $\bGamma_i$ felt by particle $i$ can include the effect of an external field. In  general, the interaction energy $V(\br_i-\br_j,\bp_i,\bp_j)$ between particles $i$ and $j$ depends on the distance between these particles and on the orientation of the dipoles they carry, as is the case in the well-known dipole-dipole interaction $V(\br_i-\br_j,\bp_i,\bp_j)=\frac{1}{4\pi\eps_0}\left[\frac{\bp_i\cdot\bp_j}{r_{ij}^3}-3\frac{(\bp_i\cdot\br_{ij})(\bp_j\cdot\br_{ij})}{r_{ij}^5}\right]$. Both the force ${\bf F}_i$ and the torque then
derive from the total potential energy $E_{\rm pot}=\frac 12 \sum_{i\neq j}V(\br_i-\br_j,\bp_i,\bp_j)$ according to
\begin{eqnarray}
{\bf F}_i &=&-\frac{\p E_\text{pot}}{\p\br_i} 
\; , \\
 {\bGamma}_i &=& -\bp_i\times \frac{\p E_\text{pot}}{\p\bp_i}
 \; . 
\end{eqnarray}
The combination $\bGamma_i\times\bp_i$ can also be written in the form
\begin{eqnarray}
\bGamma_i\times\bp_i &=& {\bf E}_i p_i^2-(\bp_i\cdot{\bf E}_i)\bp_i 
\; ,
\\
{\bf E}_i &=& -\frac{\p E_\text{pot}}{\p\bp_i}
\; . 
\end{eqnarray}
The individual dynamics of each particle being now given, we set out to determine the evolution dynamics of the local particle and dipole density.

\section{Collective modes}

The position-dipole density $\rho$ is defined by
\begin{equation}
\rho(\bx,\bpi,t)=\sum_{i}\delta(\bx-\br_i)\delta(\bpi-\bp_i)
\end{equation}
where the $\delta$'s are actually vectorial $\delta^{(3)}$'s. Our goal is to find an evolution equation for the fluctuating microscopic density $\rho$. We begin by writing that
\begin{equation}\label{deriv-1}
\begin{split}
\p_t\rho=&-\p_{\bx}\cdot\sum_i\frac{\dd \br_i}{\dd t}\delta(\bx-\br_i)\delta(\bpi-\bp_i)\\
&
 -\p_{\bpi}\cdot\sum_i\frac{\dd \bp_i}{\dd t}\delta(\bx-\br_i)\delta(\bpi-\bp_i)
\; .
\end{split}
\end{equation}
We have applied the usual rules of differential calculus, as allowed by the use of the Stratonovich discretization prescription. Note that even though \eqref{Langevin} is discretization-independent, the resulting multiplicative noise \eqref{deriv-1} is understood in the Stratonovich sense.
Then we insert in \eqref{deriv-1} the evolution equations for the individual position and dipole $\br_i$ and $\bp_i$, as given in
Eqs.~\eqref{Langevin} and \eqref{Langevin1}. This leads to
\begin{equation}\label{deriv-2}
\begin{split}
& \p_t\rho=-\p_{\bx}\cdot\sum_i\frac{1}{\gamma}\left({\bf F}_i+\beeta_i\right)\delta(\bx-\br_i)\delta(\bpi-\bp_i)\\
&\;\;\; -\p_{\bpi}\cdot\sum_i\frac{1}{\zeta} { [(\bGamma_i+\blambda_i) \times\bp_i] } \delta(\bx-\br_i)\delta(\bpi-\bp_i)
.
\end{split}
\end{equation}
Our method follows the one described in Van Kampen~\cite{vankampen2007spp}: we determine the Kramers-Moyal coefficients of $\rho$ and this leads us to identify an evolution equation (which will be of Langevin form) characterized by the same Kramers-Moyal coefficients. This path is perhaps a more physical alternative to the It\^o calculus used by Dean~\cite{0305-4470-29-24-001}. The idea is to evaluate the moments of $\Delta\rho=\rho(\bx,\bpi,t+\Delta t)-\rho(\bx,\bpi,t)=\int_t^{t+\Delta t}\dd\tau \ \p_\tau \rho$ for a random process taking place between $t$ and $t+\Delta t$, with fixed value of the density $\rho$ at the initial time $t$, and with $\p_\tau\rho$ being given by the rhs of \eqref{deriv-2}. We are interested in
\begin{equation}
\lim_{\Delta t\to 0}\frac{\langle\Delta\rho^k\rangle}{\Delta t}
\end{equation}
where, for $k\geq 2$, the $\Delta\rho$'s are evaluated at distinct $\bx,\bpi$ arguments. As it is obvious from Eq.~\eqref{deriv-2}, this limit vanishes
for any value of $k\geq 3$. This means that $\rho$ evolves according to a Langevin equation. It suffices to determine the first two nontrivial moments of
$\Delta\rho$.
The second one, which is insensitive to discretization choices, and with obvious notational shortcut, reads
\begin{equation}\label{kramersmoyal1}\begin{split}
\frac{\langle\Delta\rho\Delta\rho'\rangle}{\Delta t}=&\left[
\frac{2T}{\gamma}\p_{\bx}\p_{\bx'}\right.\\
&+\left.\frac{2T}{\zeta}\p_{\pi^\alpha}\p_{\pi^{\prime\beta}}
(\bpi\cdot\bpi'\delta^{\alpha\beta}-\pi^\alpha\pi^{\prime\beta})\right]\\&\times\rho(\bx,\bpi)\delta(\bx-\bx')\delta(\bpi-\bpi')
\; .
\end{split}\end{equation}
(We set $k_B=1$.)
The first moment is somewhat more delicate to evaluate, because of the chosen mid-point discretization scheme. Defining 
$\Pi^{\alpha\beta}=\pi^2\delta^{\alpha\beta}-\pi^\alpha\pi^\beta$, we have
\begin{equation}\label{kramersmoyal2}
\begin{split}
&
\frac{\langle\Delta\rho\rangle}{\Delta t}=
T {\gamma^{-1}} \p_{\bx}^2\rho
 + T {\zeta^{-1}} \p_{\pi^\alpha}\p_{\pi^\beta}\left(
{
\Pi^{\alpha\beta}\rho
}
\right)\\
&  \qquad \qquad + 2T  {\zeta^{-1}} \p_{\pi^\alpha}(\pi^\alpha\rho)
 - \gamma^{-1} \p_\bx\cdot\left
(\rho{\bf f}\right)
 \\
 &
 \qquad \qquad 
 - \zeta^{-1} \p_{\bpi}\cdot\left[\rho{\bf g}\times\bpi\right]
\; ,
\end{split}
\end{equation}
where
${\bf f}=-\int_{\bx',\bpi'}\p_\bx V(\bx-\bx',\bpi,\bpi')\rho(\bx',\bpi')$ and ${\bf g}=\bpi\times{\bf E}$, where the local electric field is ${\bf E}=-\int_{\bx',\bpi'}\p_{\bpi} V(\bx-\bx',\bpi,\bpi')\rho(\bx',\bpi')$. The vector fields $\bf f$ and $\bf g$ are the force and torque density, respectively.
The last two terms in \eqref{kramersmoyal2} are the direct consequence of the deterministic contributions ${\bf F}_i$ and $\bGamma_i$ in Eq.~\eqref{Langevin}.
The first three contributions are often gathered under the {\it spurious drift} terminology, in spite of their physically obvious meaning (these are diffusion terms). 

We can immediately write down the corresponding Langevin equation governing the evolution of $\rho$, that we choose to write in the It\^o scheme. 
In this discretisation scheme $\langle \Delta\rho \rangle/\Delta t$ yields the `force' term, while $\langle \Delta \rho \Delta \rho'\rangle/\Delta t$ determines the 
factor that multiplies the noise.
We cast the ensuing Langevin equation in the form of a generalized continuity equation:
\begin{equation}\label{Langevincurrent}
\p_t\rho=-\p_\bx\cdot {\bf j}-\p_{\bpi}\cdot {\bf k}
\end{equation}
where the spatial $\bf j$ current reads
\begin{equation}\label{defj}
{\bf j}=-T {\gamma^{-1}} \p_\bx\rho +
\gamma^{-1}\rho {\bf f}
+{\boldsymbol \sigma}_{\bf x}
\end{equation}
with $\langle {\boldsymbol \sigma}_{\bf x}(\bx, \bpi, t) \rangle =0$ and
\begin{eqnarray}
&& \langle\sigma_{\bf x}^\alpha(\bx,\bpi,t)\sigma_{\bf x}^\beta(\bx',\bpi',t') \rangle
\nonumber\\
&&
\qquad
=2T {\gamma^{-1}} \rho\delta^{\alpha\beta}\delta(t-t')\delta(\bx-\bx')\delta(\bpi-\bpi')
\; .
\end{eqnarray}
The dipole rotational current ${\bf k}$ has the expression
\begin{equation}\label{defk}\begin{split}
k^\alpha=&-T {\zeta^{-1}} \p_{\pi^\beta} 
\left(\Pi^{\alpha \beta}\rho\right)
\\
&
+\zeta^{-1}
\left(E_\beta \Pi^{\alpha \beta}-2T\pi^\alpha\right)\rho
+\chi^\alpha\\
=&-T {\zeta^{-1}} \Pi^{\alpha\beta}\p_{\pi^\beta} 
\rho
+\zeta^{-1} E_\beta \Pi^{\alpha \beta}\rho
+\sigma_{\boldsymbol \pi}^\alpha
\end{split}\end{equation}
and the Gaussian noise $\sigma_{\boldsymbol \pi}^\alpha$ has zero average and correlations
\begin{equation}\begin{split}
& \langle\sigma_{\boldsymbol \pi}^\alpha(\bx,\bpi,t)\sigma_{\boldsymbol \pi}^\beta(\bx',\bpi',t')\rangle
\\
&
\qquad
=
 2T {\zeta^{-1}} \,
\Pi^{\alpha \beta}
\rho \, \delta(t-t')\delta(\bx-\bx')\delta(\bpi-\bpi')
\; .
\end{split}\end{equation}
The local field ${\bf E}$ was defined after \eqref{kramersmoyal2}. There is a compact way of rewriting the Langevin
equation for $\rho$ that takes us one step closer to recent works exploiting a free-energy functional, {\it e.g.} \cite{doi:10.1080/00268976.2011.609145},
that we denote by 
\begin{equation}\label{freeenergy}\begin{split}
&{\mathscr F}[\rho]=T\int_{\bx,\bpi}\;\;\;\rho(\bx,\bpi)\ln\left[\frac{\rho(\bx,\bpi)}{\rho_\text{eq}}\right]
\\
&
\qquad
+\frac{1}{2} \int_{\bx,\bx',\bpi,\bpi'} \!\!\!\!\!\!\ \rho(\bx,\bpi)V(\bx-\bx',\bpi,\bpi')\rho(\bx',\bpi')
\; ,
\end{split}\end{equation}
where $\rho_\text{eq}=\frac{\rho_0}{4\pi}$ for a space density $\rho_0$ of dipoles with no orientational order. In case an external field ${\bf E}_\text{ext}$ is 
applied, the right-hand-side of \eqref{freeenergy} must be supplemented with a 
$-\int_{\bx,\bpi}\rho(\bx,\bpi)\bpi\cdot{\bf E}_\text{ext}$ contribution. We may then recast the Langevin 
evolution of $\rho$ in the following synthetic form:
\begin{equation}
\p_t\rho=\left(
\begin{array}{cc}\p_\bx&\p_{\bpi}\end{array}
\right)\left[\rho
\left(
\begin{array}{cc}
{\gamma^{-1}}{\bf 1}&{\bf 0}\\
{\bf 0}& {\zeta^{-1}} \bPi
\end{array}
\right)
\left(
\begin{array}{c}
\p_\bx\\
\p_{\bpi}
\end{array}
\right)
\frac{\delta {\mathscr F}}{\delta\rho}+{\boldsymbol \sigma}\right]
\; .
\label{eq:dean-rot}
\end{equation}
The Gaussian noise ${\boldsymbol \sigma}=({\boldsymbol \sigma}_{\bf x}, {\boldsymbol \sigma}_{\boldsymbol \pi})$ has correlations
\begin{equation}
\begin{split}
\langle{\boldsymbol \sigma}(\bx,\bpi,t)\otimes {\boldsymbol \sigma}(\bx',\bpi',t')\rangle
\qquad\qquad\qquad\qquad
\\ 
=2T \rho\left(
\begin{array}{cc}
{\gamma^{-1}} {\bf 1}&{\bf 0}\\
{\bf 0}&{\zeta^{-1}} \bPi
\end{array}
\right)
\qquad\qquad\qquad
\\
\; \qquad\qquad
\times\delta(\bx-\bx')\delta(\bpi-\bpi')\delta(t-t')
\; .
\end{split}
\end{equation}
Note that, by construction from the Kramers-Moyal coefficients, both Langevin equations \eqref{Langevincurrent} 
[supplemented with the definition of the local fluctuating 
currents \eqref{defj} and \eqref{defk}] and \eqref{eq:dean-rot}
are to be understood with the It\^o discretization scheme. In the absence of dipolar degrees of freedom this equation boils down to the one 
in~\cite{0305-4470-29-24-001}. 

A natural technical question that arises regards the choice of Cartesian, rather than spherical, coordinates, given the latter are ideally suited to deal with vectors 
with constant norm. Our preference for Cartesian coordinates lies in the awkward properties of the Langevin noise in the right-hand-side of \eqref{Langevin1} when it is expressed in terms of 
spherical coordinates coding for the direction of the dipolar momentum: noise is multiplicative, and has nonzero average (see \cite{discretizationgilbert} for a clear exposition); hence our choice of 
manipulating Cartesian coordinates when evaluating the Kramers-Moyal coefficients. 
It is however a very simple matter to rephrase \eqref{eq:dean-rot} in terms of ${\boldsymbol \pi}/p$ which lives on the unit sphere (here $p=||\bp_i||$ denotes the individual dipole carried by each particle). We 
arrive at the following Langevin equation for $\rho(\bx,
\bpi,t)$:
\begin{equation}
\p_t\rho=\left(
\begin{array}{cc}\p_\bx&\p_{\bpi}\end{array}
\right)\left[\rho
\left(
\begin{array}{cc}
{\gamma^{-1}} {\bf 1}&{\bf 0}\\
{\bf 0}& {\zeta^{-1}} {\bf 1}
\end{array}
\right)
\left(
\begin{array}{c}
\p_\bx\\
\p_{\bpi}
\end{array}
\right)
\frac{\delta {\mathscr F}}{\delta\rho}+{\boldsymbol \sigma} \right]
\; .
\label{eq:dean-rot-unit}
\end{equation}
where now the differential operator $\p_{\bpi}$ acts on the sphere.

So far we have achieved our primary goal, namely that of projecting the dynamics of the individual dipoles onto the dynamics of their density modes, the latter evolving according to a Langevin equation with multiplicative noise. Before we discuss the consequences of our finding we wish to propose a number of comments. First, it is important to realize that the fluctuating density field $\rho$ is a highly singular object (a sum of $\delta$ peaks). This is neither a partially smoothened nor coarse-grained version of the density. As such, the Langevin evolution for $\rho$ includes all the microscopic details of the dynamics. As can be read in the existing literature, trial density functionals, whether for statics~\cite{
BagchiChandra} or dynamics~\cite{doi:10.1080/00268976.2011.609145}, always involve dressed correlations instead of the bare interaction potential, as we have in our Eq.~\eqref{freeenergy}. It has been shown~\cite{1742-5468-2008-02-P02004}, within the more conventional framework of
translational motion of colloids, that using a dressed free-energy functional could lead to spurious findings when building up approximation schemes. Our second comment is of more technical nature. It is possible to construct a Martin-Siggia-Rose-Janssen-De Dominicis (see \cite{tauber} for a comprehensive review) path-integral formulation that is  fully equivalent to the Langevin formulation. Once the path-integral is re-expressed in terms of a pair of bosonic field via a Cole-Hopf transformation (see \cite{1751-8121-41-23-235002} or chapter 9 in \cite{tauber}), the ideal gas is described by a quadratic action, and the stage is set for approximations that expand around the ideal gas (such as a dynamical version of the virial expansion). We will return to this in the final section.  Further mathematical considerations along the lines of those presented by Jack and Zimmer~\cite{jackzimmer} on Dean's equation could certainly extend to the equation that we propose here.

\section{Correlations and Collective Effects}

The results presented in the previous section may appear to be formal. It is, however,  interesting 
to explore a few doors they open up, something
we do in this section. 

Once a Langevin equation is obtained for $\rho$, be it nonlinear and with 
multiplicative noise, simple approximations can be
devised. Suppose one is interested in correlations between dipole orientations. Then the first task 
is to evaluate the two-body correlations
$C(\bx,\bpi,t;\bx',\bpi',t') \equiv \langle\rho(\bx,\bpi,t)\rho(\bx',\bpi',t')\rangle$ (with, for concreteness, $t>t'$). 
These are 
connected to three-body correlations via a  BBGKY hierarchy, the first equation of which reads
\begin{equation}\label{BBGKY}\begin{split}
&
\p_t C (\bx,\bpi,t; \bx',\bpi',t')= T {\gamma^{-1}} \p_{\bx}^2 C(\bx,\bpi,t;\bx',\bpi',t')
\\
&
\qquad
+
T {\zeta^{-1}} \p_{\bpi} [ {\mathbf \Pi}  \p_{\bpi} C(\bx,\bpi,t;\bx',\bpi',t') ] \\
&
\qquad
+ {\gamma^{-1}} \int_{\bx'',\bpi''}\p_\bx\cdot[ \p_\bx V(\bx-\bx'',\bpi,\bpi'') 
\\
& 
\qquad\qquad\qquad \;\;\; C^{(3)}(\bx,\bpi,t; \bx'',\bpi'',t; \bx',\bpi',t') ] \\
&
\qquad
+
{\zeta^{-1}} \int_{\bx'',\bpi''} \p_{\bpi}\cdot[ {\mathbf \Pi}  \p_{\bpi}V(\bx-\bx'',\bpi,\bpi'')
\\
& 
\qquad\qquad\qquad \;\;\; C^{(3)}(\bx,\bpi,t; \bx'',\bpi'',t; \bx',\bpi',t') ]
\; , 
\end{split}\end{equation}
where $C^{(3)}(\bx,\bpi,t; \bx'',\bpi'',t; \bx',\bpi',t')$ stands for 
$\langle\rho(\bx,\bpi,t)\rho(\bx'',\bpi'',t)\rho(\bx',\bpi',t')\rangle$.
If the quantity of interest is indeed the two-point function, the difficulty then lies in expressing, by means of appropriate approximations, the right hand side of \eqref{BBGKY} in terms of two-body correlations. We illustrate in the following subsections two strategies that our functional formalism allows us to phrase in an elegant way.

\subsection{The Random Phase Approximation (RPA) closure scheme}

The simplest closure scheme, which in liquid state
theory~\cite{HansenMcDonald2006} is often termed the RPA
approximation, translates, in the density language, into a Gaussian truncation for the fluctuations around the mean density. Suppose we are sitting in a
disordered phase with equilibrium density  given by 
$\rho_\text{eq}=\langle \rho\rangle=\frac{\rho_0}{4\pi}$, where $\rho_0$ is the actual particle density.
We immediately obtain that in the time translation-invariant regime, with $\tau=t-t'$, correlations evolve according to
\begin{eqnarray}
\begin{split}
&
\p_\tau C(\bx,\bpi,\bx',\bpi',\tau)=
\\
& \qquad
(
{\gamma^{-1}} \p_\bx^2+
{\zeta^{-1} \p_{\bpi} \cdot{\boldsymbol \Pi} \p_{\bpi}}
)
\Big[TC(\bx,\bpi,\bx',\bpi',\tau) 
\\
&
\quad 
+\frac{\rho_0}{4\pi }\int_{\bx'',\bpi''} \!\!\!\!\!\!\!\!\!\! V(\bx-\bx'',\bpi,\bpi'')C(\bx'',\bpi'',\bx',\bpi',\tau)\Big]
\;\;\;\;\;\;
\end{split}
\end{eqnarray}
in full agreement, of course, with the entire Section V in \cite{BagchiChandra}, but at virtually no technical cost. The RPA approximation is 
notoriously unable to address nonlinear effects that are observed in denser systems as the glass transition. The next subsection illustrates, following the ideas presented in \cite{0305-4470-38-20-L03}, 
but supplemented with the formalism of \cite{1742-5468-2008-02-P02004}, how functional methods can efficiently match other theoretical approaches.

\subsection{A memory kernel approach}

A more refined approximation scheme is the {\it mode-coupling approach}.
The mode-coupling theory (MCT) has been first applied \cite{bengtzelius_dynamics_1984} to point-like particles with Hamiltonian dynamics and it was 
later extended by Szamel and L\"owen~\cite{szamel_mode-coupling_1991} to point-like particles with overdamped Langevin dynamics. 
For particles with an internal structure, Schilling and Scheidsteger~\cite{PhysRevE.56.2932} extended the original calculation for 
Hamiltonian dynamics to linear molecules. Our purpose here is twofold. We want to illustrate the elegance and compactness of functional 
methods by showing that a quadratic approximation to $\mathscr F$ coupled to a one loop expansion lead to a mode-coupling equation for 
two-body correlations. We will thus present a result for a mode-coupling equation for polar molecules that tries to extend that of 
\cite{PhysRevE.56.2932} to the Langevin case (in much the same way as \cite{szamel_mode-coupling_1991} extended 
\cite{bengtzelius_dynamics_1984}). 

One way to understand the approximations behind the mode-coupling equation for $C$ that can be 
found in the literature~\cite{PhysRevE.56.2932} is to postulate a simple quadratic approximation for 
the equilibrium free-energy functional ${\mathscr F}$ 
that we express in terms of $\psi=\rho-\frac{\rho_0}{4\pi}$,
\begin{equation}
\beta {\mathscr F}[\psi]=\frac 12\int_{\bx,\bpi,\bx',\bpi'}
\!\!\!\!\!
\psi(\bx,\bpi) { C}_{eq}^{-1}(\bx,\bpi;\bx',\bpi')\psi(\bx',\bpi')
\end{equation}
and to endow the density modes with the same dynamics as in \eqref{eq:dean-rot}. 
The equilibrium static correlations are given by $C_{eq}$ {and the inverse has to be understood 
operationally}.  

In a dynamical functional formulation {\it \`a la} Janssen-De Dominicis~\cite{janssen_lagrangean_1976,de_dominicis_techniques_1976} the action reads
\begin{equation}\label{act-truc}
\begin{split}
& S[\bar{\psi},\psi]=
\int_{t,\bx,\bpi}
\left[
{i} \bar{\psi}\p_t\psi
+T{\gamma^{-1}} \rho \ {(} \p_\bx {i} \bar{\psi} {)} \cdot\p_\bx(C_{eq}^{-1}*\psi)
\right.
\\
&
\;\;\;
+T{\zeta^{-1}} {\rho \ (} \p_{\pi^\alpha} {i}  \bar{\psi}  {)} \Pi^{\alpha\beta}\p_{\pi^\beta}(C_{eq}^{-1} {*} \psi)
\\
&
\;\;\;
\left.
-T{\gamma^{-1}} \rho \ (\p_\bx i\bar{\psi})^2
-T{\zeta^{-1}} \rho \ {(} \p_{\pi^\alpha}i\bar{\psi} {)} \Pi^{\alpha\beta}\p_{\pi^\beta}i\bar{\psi}
\ \right]
\end{split}
\end{equation}
where $*$ is defined by
\begin{displaymath}
C_{eq}^{-1}*\psi(\bx,\bpi, t)=\int_{\bx'',\bpi''} \!\!\!\! 
C_{eq}^{-1}(\bx,\bpi;\bx'',\bpi'')\psi(\bx'',\bpi'', t)
\; . 
\end{displaymath}
The only interaction terms of the resulting field theory are cubic, and they arise from the 
multiplicative $\rho$ factor, bearing in mind that $\rho=\frac{\rho_0}{4\pi}+\psi$.  {The generating functional 
is ${\cal Z} = \int {\cal D}\psi{\cal D}\overline \psi \ P_{\rm i}(\psi(-{\cal T})) \ \ee^{-S[\psi,\overline\psi]}$ where $P_{\rm i}$ is the field probability 
 at the initial time $-{\cal T}$ that, in canonical equilibrium, is given by $P_{\rm i}(\psi(-{\cal T})) =  Z^{-1} e^{-\beta {\mathscr F}(\psi(-{\cal T}))}$, with $Z$
 the normalisation constant.}

The action $S + \ln P_{\rm i}$ (as well as the measure) are invariant under the linear transformation
{
\begin{eqnarray}
\psi(\bx, \bpi; t) &\mapsto & \psi(\bx, \bpi, -t) 
\; , 
\label{eq:Tr1}
\\
{i} \overline\psi(\bx, \bpi; t) &\mapsto& - {i} \overline\psi(\bx, \bpi, -t)  
+ \frac{1}{T} \frac{\delta {\mathscr F}}{\delta \psi(\bx,\bpi,- t)}
\; . 
\label{eq:Tr2}
\end{eqnarray} 
The stochastic equation on $\rho$, although with 
multiplicative noise and in the It\^o convention, does not imply any unconventional chain-rule for the time-derivatives of functionals of $\rho$. The corresponding dynamic action \eqref{act-truc} is also insensitive to the discretization rule.

A more compact way of writing \eqref{act-truc} is to define a differential operator $\vec{D}$ acting on the 
$3+3$-dimensional physical and dipolar spaces, defined by 
$\vec{D}=({\gamma^{-1/2}} \p_{\bx}, {\zeta^{-1/2}} i{\bf L})$ (where $i{\bf L}=\bpi\times\p_{\bpi}$), 
which allows us to rewrite the action as
\begin{equation}\label{act-dyn}\begin{split}
S[\bar{\psi},\psi]=&
\int_{t,\bx,\bpi}
 \left[ 
{i} \bar{\psi}\p_t\psi+T 
\frac{\rho_0}{4\pi}\vec{D} {i} \bar{\psi}\cdot\vec{D}(C_{eq}^{-1}*\psi)
\right.
\nonumber\\
&-T
\frac{\rho_0}{4\pi}(\vec{D}i\bar{\psi})^2\\
&
\left.
+T\psi\vec{D} {i} \bar{\psi}\cdot\vec{D}(C_{eq}^{-1}*\psi)-T\psi(\vec{D}i\bar{\psi})^2
\right]
\; . 
\end{split}
\end{equation}

Formally at least, we are working within the framework of ``class II" systems considered 
in~\cite{0305-4470-38-20-L03}, and our contribution now consists in dealing with the more complex 
space-structure and in exploiting time-reversal invariance [Eqs.~(\ref{eq:Tr1}) and (\ref{eq:Tr2})]
in a slightly more compact fashion. With standard notations, 
the Schwinger-Dyson equation for the four correlations between the fields, gathered in a matrix $G$, reads 
$G_0^{-1}G=\mathbf{1}-\Sigma G$, where $G_0$ is the propagator related to the Gaussian part of the action \eqref{act-dyn}. 
The matrix $G$'s entries are the various $\bar{\psi}\bar{\psi}$, $\bar{\psi}\psi$ or $\psi\psi$ correlations. The former being zero by causality, 
$G$ can be expressed solely in terms of $C=\langle\psi\psi\rangle$ and $R=\langle {i} \bar{\psi}\psi\rangle$. 
{Assuming time-translational invariance}
we project the Schwinger-Dyson equation for each matrix entry, to arrive at 
\begin{equation}\begin{split}
& \p_t C(t) -T\frac{\rho_0}{4\pi}\vec{D}^2(C_{eq}^{-1}*C(t))
\qquad\qquad\qquad
\nonumber\\
&\qquad = -\int_{-\infty}^{+\infty}\dd\tau \ \Sigma_{\bar{\psi}\bar{\psi}}(t-\tau) { \ * \ }R(-\tau)
\\
&\qquad\;\;\;\; -\int_{-\infty}^{+\infty}\dd\tau \ \Sigma_{\bar\psi \psi}(t-\tau) {\ * \ } C(\tau)
\end{split}
\end{equation}
To alleviate the overall appearance of the equation we have momentarily dropped the space and orientational coordinates
and we have only kept the time-dependence. A similar equation holds for the evolution of $R$ but, under the assumptions we are using,
it is actually strictly redundant. 

The time-reversibility symmetry (\ref{eq:Tr1}) and (\ref{eq:Tr2})
leads to a 
tremendous simplification, because it is linear and it leaves the Gaussian part and the interaction part of the dynamic action 
separately invariant. Hence, on top of being causal, 
$R$ is simply related to $C$. Using the symmetry \eqref{eq:Tr1} and \eqref{eq:Tr2} we recover the result found by Deker and Haake~\cite{deker_fluctuation-dissipation_1975} in a straightforward fashion; given that $\langle i\bar{\psi}(0)\psi(t)\rangle=\langle\left(- {i} \overline\psi( 0)  
+ \frac{1}{T} \frac{\delta {\mathscr F}}{\delta \psi( 0)}\right)\psi(-t)\rangle$, we arrive at
\begin{equation}
R(t)+R(-t)=C_{eq}^{-1}*C(t)
\; . 
\end{equation}
Using the techniques developed in \cite{1742-5468-2008-02-P02004}, a similar relation can be shown to hold for the vertex functions
 \begin{equation}
 \Sigma_{\bar{\psi}\psi}(t)+\Sigma_{\bar{\psi}\psi}(-t)=-C_{eq}^{-1}*\Sigma_{\bar{\psi}\bar{\psi}}(t)
 \; . 
 \end{equation} 
Using these proportionality relations along with the causality of $R$ and $\Sigma_{\bar{\psi}\psi}$, we immediately arrive at
\begin{equation}
\label{exactmct}
\begin{split}
& \p_t C(t) -T\frac{\rho_0}{4\pi}\vec{D}^2(C_{eq}^{-1}*C(t))
\qquad
\\
& 
\qquad =-\int_{0}^{t}\dd\tau \ \Sigma_{\bar\psi\psi}(t-\tau) { \ * \ } C(\tau)
\end{split}
\end{equation}
which is still devoid of any approximation (beyond the quadratic assumption on $\mathscr F$ and equilibrium). 

The only task left is to evaluate the vertex function $\Sigma_{\bar{\psi}\psi}$, which can be done in a loop expansion or, 
in other words, in powers of the cubic interaction terms. If $\Sigma_{\bar{\psi}\psi}$ is determined in a perturbation 
theory then, to leading order, both \eqref{exactmct} and 
\begin{equation}\label{approxmct}
\begin{split}
& \p_t C(t) -T\frac{\rho_0}{4\pi}\vec{D}^2(C_{eq}^{-1}*C(t) )
\qquad
\\
& 
\quad
=
\frac{4\pi}{\rho_0 T}\int_{0}^{t}\dd\tau \ \Sigma_{\bar{\psi}\psi}(t-\tau) { \ * \  [} C_{eq}*(\vec{D}^2)^{-1}\p_\tau C(\tau) {]}
\end{split}\end{equation}
are equivalent. Equation \eqref{approxmct} has the exact same structure as the mode-coupling equation derived 
by~\cite{szamel_mode-coupling_1991} 
on condition that the memory kernel can be expressed as a quadratic form in $C$.\begin{figure}[h]
\includegraphics[width=7cm]{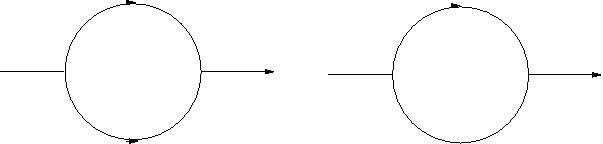}
\caption{The left diagram has two $R$ internal lines, while the right one has an $R$ and a $C$ internal line
(an arrowed leg denotes a response field $\bar{\psi}$ and a plain leg denotes the density field $\psi$). 
}
\label{figure:twomctdiagrams}
\end{figure}
This is indeed the case because the two diagrams shown in Fig.~\ref{figure:twomctdiagrams} that contribute 
to $ \Sigma_{\bar{\psi}\psi}$ have internal loops involving $C$ and $R$.
 
The algebra simplifies greatly if we focus on the Fourier transform of $C(\bx,\bpi;\bx',\bpi',t)$, 
which we denote by $C(\bk;m,l;m',\ell',t)$ (where $\bk$ is conjugate to $\bx-\bx'$, and  $\ell,m$, $\ell',m'$ are the indices of the 
coefficients of the expansion of $C$ on the spherical harmonics). A bit of tedious algebra leads to the one-loop expression for $ \Sigma_{\bar{\psi}\psi}$ 
in terms of $C$. To indicate space and orientational degrees of freedom we use the short notation $n=(\bx_n,\bpi_n)$ and use the Einstein convention on repeated indices. The memory kernel $ \Sigma_{\bar{\psi}\psi}(1,2);t)$ has the following explicit expression:
\begin{equation}\begin{split}
& \Sigma_{\bar{\psi}\psi}(1,2;t)
=T^2D_2^{\beta}
\nonumber\\
&
\;\; \times \left[-2(D_1^\alpha R(2,1)) (D_2^{\beta}C_{eq}^{-1}(2,3))(D_1^\alpha R(3,1))\right.\\
& \quad
+(D_1^\alpha R(4,1))(D_1^\alpha C_{eq}^{-1}(1,3))(D_2^\beta C_{eq}^{-1}(2,4))C(2,3)
\\
&
\quad
+(D_1^\alpha R(2,1))(D_1^\alpha C_{eq}^{-1}(1,3))(D_2^\beta C_{eq}^{-1}(2,4))C(3,4)
\\
&
\quad
+(D_3^\alpha C_{eq}^{-1}(3,1))(D_2^\beta C_{eq}^{-1}(2,4)) (D_3^\alpha R(4,3))C(3,2)\\&\quad
+\left.(D_3^\alpha C_{eq}^{-1}(3,1))(D_2^\beta C_{eq}^{-1}(2,4)) (D_3^\beta R(2,3))C(3,4)\right]
\end{split}
\end{equation}
where $\vec D_n$ means that the derivatives are taken with respect to ${\mathbf x}_n$ and ${\boldsymbol \pi}_n$. While the above expression for the memory kernel $\Sigma_{\bar{\psi}\psi}$ appears overly intricate, owing to the absence of 
translation invariance in dipolar space, a simple check that this result must be correct is first to replace 
$\vec{D}$ with $\p_\bx$ while forgetting about dipolar coordinates, and then go to Fourier space. The standard mode-coupling memory kernel found  
in \cite{bengtzelius_dynamics_1984,szamel_mode-coupling_1991} is then immediately recovered, on condition that the inverse structure factor $C_{eq}^{-1}$  is replaced with the direct correlation function$-C_{eq}^{-1}+1/(\rho_0/4\pi))$. This proviso comes from our using a Gaussian truncated equilibrium energy even for the ideal gas, instead of the standard entropic $\rho\ln\rho$ contribution arising from Poisson statistics. It was shown explicitly in \cite{PhysRevE.89.012150} (section IV.A) how to recover the  original mode-coupling kernel with the direct correlation function appearing in the memory kernel, instead of the structure factor. With this caveat in mind, keeping dipolar coordinates and Fourier 
transforming real space, we realize that our memory kernel is identical to that presented in \cite{PhysRevE.56.2932}, and which was derived there for Hamiltonian dynamics by means of more standard projection techniques.

\section{Conclusions}

We have presented a derivation of the Dean-Kawasaki equation for systems with translational and rotational degrees of freedom. After having written this
equation, we have discussed the BBGKY hierarchy of equations ruling the evolution of the correlation functions and two approximations, 
the RPA Gaussian closure and the MCT.

One can envisage to extend the renormalized perturbation theory techniques proposed by Kim and Kawasaki for the
field theory associated to the density field of point-like particles~~\cite{1751-8121-40-1-F04, 1742-5468-2008-02-P02004},
to the present case with both translational and rotational degrees of freedom. In this way one should be able to
obtain evolution equations for the density correlation functions probably differing from those obtained in \cite{PhysRevE.62.5173}. This is of great interest since the onset of glassiness does not occur at the same temperatures for translational degrees of freedom and for rotational ones. Hence a global treatment of both translational and rotational degrees of freedom would allow one to explicitly disentangle the former from the latter. 
Applications to other fields can also be foreseen.

On the one hand, the interplay between translational and rotational motion is of particular interest in active matter
as active units are typically elongated. This is the case in both natural, as in bacteria, or artificial, as in chemically propelled
nanorod, systems. Activation can be modeled at the level of the departing Langevin equation for the single units, as
done in~\cite{Wensik-etal,Suma14} for dumbbell particles, and the extension to Eq.~(\ref{eq:dean-rot}) be readily worked out.
One could also choose a simpler approach and add sources of activation at the level of Eq.~(\ref{eq:dean-rot}) by
supplementing it with non thermal sources of noise or by embedding it  in an external density dependent velocity field.


Multiplicative noise always arises in mechanical overdamped rotational Brownian motion, which in turn is in use for handling the thermal random rotational dynamics of permanent electric dipoles, as well as Langevin paramagnets relevant to the Debye relaxation mechanism in ferrofluids~\cite{raikher-shliomis,doi:10.1142/9789814355674}. Multiplicative noise is also important for handling the N\'eel relaxation mechanism pertaining to the thermally activated magnetization reversal {\it inside} magnetic nanoparticles, where the micromagnetic magnetization rotation mode is the Stoner-Wohlfarth uniform mode~\cite{Aharoni-Ferro}. In particular, as shown in App.~\ref{app:magnetic}, our formalism allows the direct treatment of the dynamics of interacting magnetic nanoparticle assemblies, had the magnetic particles been frozen in a solid matrix -in which case only the spin degrees of 
freedom are relevant- or with coupled mechanical and spin degrees of freedom if the assembly consists of colloidal suspensions in a liquid carrier. The problem of handling such dynamics is extremely important both from the fundamental and applied points of view~\cite{dorman-fiorani-troncACP,0022-3727-42-1-013001}.


\vspace{1cm}

\acknowledgements

We warmly acknowledge discussions with Bongsoo Kim. While this work was completed G. S. Lozano benefitted from an Alicia Moreau guest professorship at the
Universit\'e Paris Diderot. We acknowledge financial support from PICT-2012-0172 (Argentina), PIP CONICET 2012 0931, and the ECOS-Sud A14E01 collaboration. 
LFC and FvW are members of Institut Universitaire de France.

\appendix

\section{Magnetic dipoles}
\label{app:magnetic}

In this appendix we briefly review the magnetic analog of the calculation presented in the body of this paper pointing to existing differences with electric dipoles. The physical system we have in mind is a suspension of anisotropic magnetica nanoparticles, which, on top of being characterized by a position $\br_i$ and a direction $\bpi_i$ (which does not have to be related to an actual electric dipole and may just be a director in case the particle is anisotropic), now bear a magnetic dipole $\bmu_i$. In addition to the individual evolution equations \eqref{eqdebase} and \eqref{eqdebase1}, each magnetical dipole evolves according to the Gilbert-Landau-Lifschitz equation \cite{gilbert1955,brown1963} (and the notations of \cite{discretizationgilbert}):
\begin{equation}
\frac{\dd\bmu_i}{\dd t}=-\gamma_0\bmu_i\times\left(-\p_{\bmu_i} E_{\rm pot}+\bh_i-\frac{\eta}{\mu_s} \frac{\dd\bmu_i}{\dd t} \right)
\label{startmag0}
\end{equation}
where $\bh_i(t)$ is a Gaussian random white noise with variance $\frac{2T\eta}{\mu_s}$ 
The potential energy  $E_{\rm pot}$ may now depend on the $\bmu_j$'s. The latter equation, using the notation ${\bf H}_i=
-\p_{\bmu_i} E_{\rm pot}$, is equivalent to
\begin{eqnarray}
\frac{\dd\bmu_i}{\dd t}
&=&
-\frac{\gamma_0}{1+ \gamma_0^2\eta^2}\bmu_i
\nonumber\\
&& 
\times 
\left(\bH_i+\bh_i+\frac{\eta\gamma_0}{\mu_s} \bmu_i\times [\bH_i+\bh_i]\right)
\; . 
\label{startmag}
\end{eqnarray}
The Langevin equations \eqref{startmag0} and \eqref{startmag} are to be understood in the Stratonovich discretization scheme that 
ensures the conservation of the modulus of each magnetic moment. 

The local density 
$\rho(\bx,\bpi,\bmm,t)=\sum_i\delta(\bx-\br_i)\delta(\bpi-\bpi_i)\delta(\bmm-\bmu_i)$, as in the body of the paper, will also evolve according to a Langevin equation 
with the Ito discretisation prescription that reads
\begin{equation}\label{Deanextended}
\p_t\rho=\left(
\begin{array}{ccc}\p_\bx&\p_{\bpi}&\p_{\bmm}\end{array}
\right)\left[\rho
\left(
\begin{array}{ccc}
\frac{{\bf 1}}{\gamma}&{\bf 0}&{\bf 0}\\
{\bf 0}&\frac{\bPi}{\zeta}&{\bf 0}\\
{\bf 0}&{\bf 0}& \frac{{\eta}}{{\mu_s}}{\bf M}
\end{array}
\right)
\left(
\begin{array}{c}
\p_\bx\\
\p_{\bpi}\\
\p_{\bmm}
\end{array}
\right)
\frac{\delta {\mathscr F}}{\delta\rho}+{\boldsymbol\sigma}
\right]
\; 
\\
\end{equation}
where the new tensor ${\bf M}$ has the expression 
\begin{eqnarray*}
M^{\alpha\beta}  
=\left(\frac{\gamma_0}{1+\gamma_0^2\eta^2}\right)^2 
\!\!
\left(1+\frac{\eta^2\gamma_0^2}{\mu_s^2}\bmm^2\right)
\left({\bmm}^2\delta^{\alpha\beta}-m^\alpha m^\beta\right)
\; . 
\end{eqnarray*}

The noise ${\boldsymbol \sigma}$ appearing in the right-hand side of the first-order differential 
equation on $\rho$ not only has three components for spatial diffusion and three components for the diffusion of the director, 
but also another set of three components for diffusion in magnetization space, ${\boldsymbol \sigma} = ({\boldsymbol \sigma}_{\bf x}, 
{\boldsymbol \sigma}_{\boldsymbol \pi}, {\boldsymbol \sigma}_{\bf m})$. Its correlations, in the magnetization subspace, are given by
\begin{displaymath}
\begin{split}
& \langle {\boldsymbol \sigma}_{\bf m}(\bx,\bpi,\bmm,t)\otimes {\boldsymbol \sigma}_{\bf m}(\bx',\bpi',\bmm',{t'})\rangle=
\qquad\qquad\qquad\qquad
\\
& 
\qquad
\frac{2 T {\eta}}{\mu_s} {\bf M} \ \rho(\bx,\bpi,\bmm,t)\delta(\bx-\bx')\delta(\bpi-\bpi')\delta(\bmm-\bmm')
\\
&
\qquad
{\delta(t-t')}
\; . 
\end{split}
\end{displaymath}

The functional $\mathscr F$ involved here is a direct extension of the 
one considered previously, and incorporates the magnetization degrees of freedom and their interactions; beyond the standard dipole-dipole interaction, such contributions as ${\mathscr F}=-K\sum_i (\bp_i\cdot\bmu_i)^2=-K\int_{\bx,\bpi,\bmm}(\bpi\cdot\bmm)^2\rho$ that favor the alignement of the magnetic dipole along the easy axis of the particle may be included.

In the $\eta\gamma_0\gg 1$ limit which is relevant to ferrofluids~\cite{raikher-shliomis}, ${\bf M}$ simplifies into $M^{\alpha\beta}=(\mu_s\eta)^{-2}
\ {{\bmm}^2} \ 
\left({\bmm}^2\delta^{\alpha\beta}-m^\alpha m^\beta\right)$.
Note that a special form of \eqref{Deanextended} has implicitly been used by D\'ejardin~\cite{dejardin-magnetic-relax} who used a mean field version of this equation in order to handle the 
N\'eel magnetization reversal in assemblies of weakly interacting uniaxial magnetic nanoparticles, with the result that the thermally activated reversal time is consistent with a 
Vogel-Fulcher behavior, as was first expected by Shtrikman and Wohlfarth~\cite{Shtrikman1981467} on qualitative grounds, and which is valid for all values of the dissipation, 
for uniaxial particles within  the mean field approximation.

\bibliography{rotational-bibliography}


\end{document}